\documentclass[]{spie}  %>>> use for US letter paper
\usepackage{amsmath,amsfonts,amssymb}
\usepackage{graphicx}
\usepackage[colorlinks=true, allcolors=blue]{hyperref}
\usepackage[export]{adjustbox}
\usepackage{siunitx}

\title{HARMONI at ELT: line spread functions in a diffraction limited spectrometer}

\author[a]{Stephen P. Todd}
\author[a]{{\'E}amonn J. Harvey}
\author[a]{Anna MacIver}
\author[a]{William Taylor}
\author[b]{Eduard Muslimov}
\author[c]{Kjetil Dohlen}
\author[b]{Matthias Tecza}
\author[d]{Magali Loupias}
\author[e]{Paula Ba{\~n}ares-Palacios}
\author[f]{Mark Swinbank}
\author[b]{Ryan Griffiths}
\affil[a]{UKATC, Royal Observatory Edinburgh, Blackford Hill, Edinburgh, UK}
\affil[b]{Oxford University, Denys Wilkinson Building, Keble Road, Oxford OX1 3RH, UK}
\affil[c]{Aix-Marseille Université, CNRS, CNES, LAM, Marseille, France}
\affil[d]{Universit{\'e} Lyon 1, CNRS, Centre de Recherche Astrophysique de Lyon (CRAL), Saint~Genis-Laval, France; }
\affil[e]{Instituto de Astrofísica de Canarias (IAC), Tenerife, Spain}
\affil[f]{Department of Physics, Durham University, South Road, Durham, UK}

\authorinfo{Correspondence to SPT. Email: stephen.todd@stfc.ac.uk}

\begin{document} 
\maketitle

\begin{abstract}
HARMONI is the first light, adaptive optics assisted, near-IR integral field spectrograph for the ELT. It covers a spectral range from 800~nm to 2450~nm with resolving powers from 3000 to 7000 and spatial sampling of 25~mas and 6~mas. It can operate in two adaptive optics modes - SCAO (including a high contrast capability) and MCAO. The project is resuming its final design phase after a rescope design phase in 2025. 

Diffraction of the pupil becomes significant in a spectrograph where the slit width is comparable to the diffraction limited PSF. When the spatial coherence due to the narrow slit is considered, the resulting line spread function can be narrower than the geometric width of the input slit, with a non-linear dependence on the size of the pupil aperture after the slit. We outline the impact of these diffraction and spatial filtering effects on the line spread function of HARMONI and identify parameters that should be considered when designing a diffraction limited spectrograph.
\end{abstract}


\section{INTRODUCTION}

When designing an astronomical spectrometer it is important to ensure that the spectral lines are adequately sampled by the detector pixels. The exact requirement will vary depending on the application, but generally it is desirable to match the full width half maximum (FWHM) of the line profile to 2 -- 2.5 detector pixels. Achieving 2 pixel sampling is often described as Nyquist sampling. This is not strictly accurate - the information present in higher spatial frequencies within the spectrum will depend on the shape of the line profile - but is generally accepted as a sensible rule of thumb. 

In the early stages of the design process it is common practice to assume that the FWHM will be given by the geometric width of the slit image, potentially broadened by some optical aberrations. Historically, spatial image quality has often not been a priority for spectrometers, which generally operated in the seeing limited regime. The development of large format infrared detectors and image slicing integral field units has led to development of a class of instruments where both spectral and spatial resolution are important. Combining these with adaptive optics corrected telescopes such as the ESO ELT (or space telescopes) brings us into a different regime, where the slit or slice width is designed to sample the diffraction limited resolution of the telescope. In this regime the effects of diffraction and spatial coherence create spatial filtering effects due to clipping of the diffracted pupil image at the grating aperture. This produces slit images on the detector which are significantly narrower than would be expected from geometric optics alone. These effects have long been known - outlined by Wadsworth in 1897 \cite{Wadsworth01051897} - but are often not considered in the initial design process. A number of instruments operating in the diffraction limited regime have observed narrower than expected lines due, at least in part, to the effects that we discuss here \cite{Davies2023}, \cite{Dionne2008}.

Analytic expressions have been derived for the slit image profiles for an ideal spectrograph when these diffraction effects are taken into account \cite{Casini:14}\cite{Mielenz:67}. These papers are written from the perspective of a laboratory spectroscope, and quantify the line width based on the Rayleigh criterion. Here we relate these formulae directly to the top level design parameters of an astronomical spectrograph and measure the line widths based on the full width at half maximum (FWHM). Using this approach we have identified an issue of expected undersampling in the HARMONI spectrograph. While the project is already well advanced, it is still in the design phase, so we outline the proposed mitigation strategy to bring the sampling close to 2 pixels in all operating modes.

\section{Overview of HARMONI}

HARMONI is an integral field spectrograph for the ELT, operating over a wavelength range of 0.8--\SI{2.45}{\um}. It is designed to take full advantage of the adaptive optics corrected image quality of the ELT, sampling the image with 6 mas spatial pixels (spaxels), while also allowing integral field spectroscopy over a larger field of view with 25 mas sampling. A rescope exercise has recently been carried out to simplify the HARMONI design \cite{Dunlop2026}\cite{Maciver2026}. The number of sampling scales and wavelength bands has been reduced. The spectrograph optics have also been substantially redesigned, moving to an all-reflective freeform based design \cite{Muslimov2026}. 

\begin{figure}
\includegraphics[width=15cm, center]{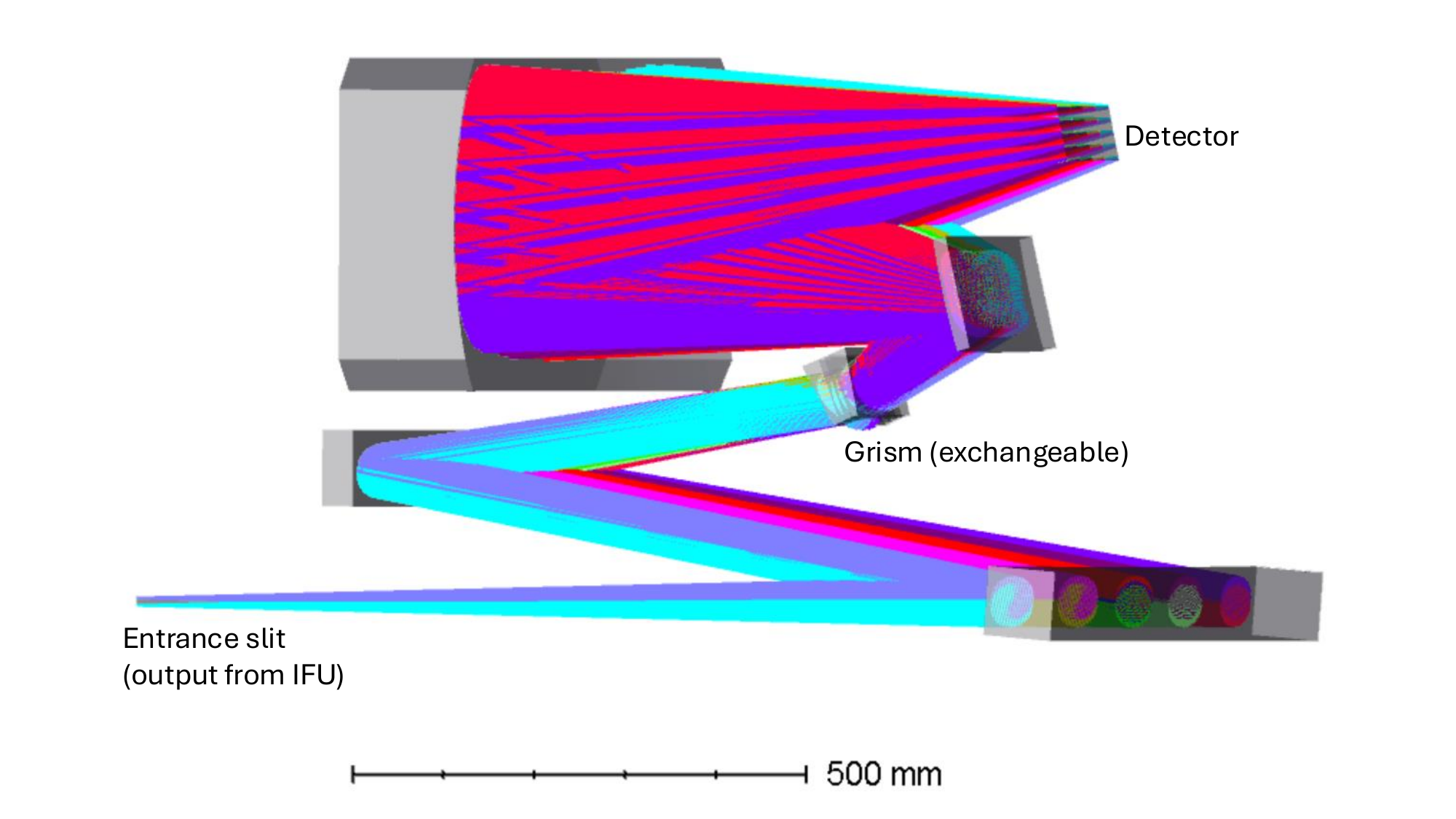}
\caption{The layout of the HARMONI IFS Spectrograph (ISP) following the rescope process. The collimator and camera are each formed of two freeform reflective surfaces with a set of grisms used as dispersing elements. More details of the design can be found in Muslimov et.\ al. \cite{Muslimov2026}.}
\label{fig:ISP_layout}
\end{figure}

Light from the ELT is relayed to the HARMONI entrance focal plane by MORFEO \cite{Ciliegi2024}. The HARMONI science field is split into eight sub-fields, each passing to an image slicing integral field unit to provide spectra over a single contiguous field measuring 204x152 spaxels (“SPAtial piXELS”) in size. Two spaxel scales of 25 x 25 mas and 6 x 6 mas per spaxel give fields of view on the sky of . Spectral resolutions of ~3000 and ~7000 are provided in both spatial scales, covering the wavelength range (not simultaneous) of 0.8–2.45 micrometres. This is a complex optical system made up of many separate subsystems as shown schematically in Figure 1. In this paper we consider only the components and subsystems that make up the ‘Science Path’ of HARMONI – the optical path from the focal plane delivered by the telescope to the science detectors. This path includes a number of sub-systems: 
\begin{itemize}
   
\item IFS Pre-Optics (IPO) – includes a collimator to image the telescope pupil onto an exchangeable cold stop or pupil mask, and two separate sets of relay optics which can be selected to provide the required spaxel scale. The IPO introduces an anamorphic magnification factor of 2 to give equal sampling in both dimensions while also imaging the slice width of the IFU onto two detector pixels for spectral sampling (based on geometric imaging).  

\item Integral Field Unit (IFU) – the image from the IPO is projected onto a field splitter mirror which cuts the science field into 8 rectangular sub-fields. Each of these sub-fields is magnified onto a slicing mirror made up of 38 slices. The output slit images from these 8 slicing mirrors are arranged in pairs, with each pair placed end to end to form a single slit image.
\item
IFS Spectrograph (ISP) – four identical spectrographs, each receiving the output slit produced from two slicing mirrors of the IFU. The spectrographs use all reflective freeform optics to image the 
spectra onto a focal plane consisting of two Teledyne HAWAII-4RG detectors. Dispersion is provided by 
transmissive VPH gratings. Multiple gratings are used to cover range of wavelength settings.
\end{itemize}

The sampling of the spectral line spread function (LSF) impacts both calibration of the spectra, and the science measurements.  If the sampling is too low, narrow spectral features will be undersampled, resulting in uncertainties in their measured centroids (even if the shape of the line profile is well constrained).  In the calibration, this results in systematic errors in the wavelength calibration, which propagate directly to the science analysis (e.g. when trying to measure the absorption or emission line centroid of narrow features from an astrophysical source, distinguish closely separated lines, or when trying to measure the shape of a spectral feature).

Furthermore, in the near-infrared the sky OH emission must be subtracted, and narrow telluric water absorption features corrected for.  If an instrument suffers even the smallest amount of spectral flexure between exposures, the sky subtraction and telluric correction can leave significant systematic residuals.  These systematics are more significant when the spectra are undersampled.  The systematics then have significant impacting the science measurements.

Oversampling is also undesirable, coming at the cost of both wavelength coverage (for a fixed detector size), and readout noise.  Typically for astronomical purposes, sampling at two pixels per FWHM represents an acceptable trade-off that provides good wavelength calibration, with centroids typically to $<0.1$ spectral pixels (with some dependence on signal-to-noise, line phasing and morphology, e.g. see Robertson 2017 \cite{Robertson2017}).  Sampling with a minimum of two spectral pixels is usually therefore mandated by science to provide the acceptable trade off between calibration and science analysis, so long as the spectral flexure of the instrument is under control.

\begin{figure}[h]
\includegraphics[width=\textwidth]{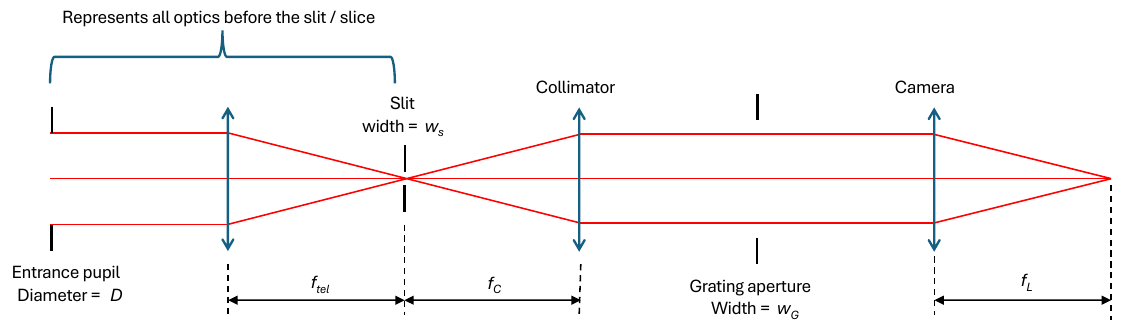}
\caption{A highly simplified model of a spectrograph, based on paraxial (ideal) lenses to represent the optics before the slit plane and the collimator and camera within the spectrograph.}
\label{fig:paraxial model}
\end{figure}

\section{Modelling an idealised spectrograph}
\subsection{The importance of coherence}

A simplified model of a spectrograph is shown in figure \ref{fig:paraxial model}. The optical assemblies of a real system are represented by ideal (paraxial) lenses. This model could represent a simple slit spectrometer but could also represent a single slice of an image slicing integral field spectrograph such as HARMONI. Even when considering extended sources (such as sky emission lines) which overfill the width of the slit it is still essential to include the entrance pupil within the model. The limited aperture of the pupil introduces a degree of spatial coherence into the beam. If we consider the angular extent of the pupil as viewed from the slit then we can identify two limiting cases of spatial coherence. The illumination of the slit becomes fully coherent if the illumination comes from a small pupil aperture at a large distance from the slit and fully incoherent when the pupil image is large and close to the slit. 

Analytic expressions for the line profiles in these two extreme cases were derived by Casini\cite{Casini:14}. These are given in terms of the width of the slit and grating apertures, $w_S$ and $w_G$ and the focal lengths of the collimator and camera, $f_C$ and $f_L$. The slit length is assumed to be very large compared to the diffraction limited spot size, and the grating aperture is rectangular and oversized in the $y$ dimension. Diffraction in the $y$ dimension is neglible, allowing the line profiles to be written as a one-dimensional function of $x$. 

To simplify the notation we define the quantities
\begin{equation}
    \theta_x^{\pm}=\frac{\pi}{2}\frac{w_s/f_C \pm 2x/f_L}{\lambda/w_G}
\end{equation}
The line profiles at the output focal plane in the two cases can be expressed in terms of the sine integral function  $\text{Si}(x) = \int_0^x \frac{\sin t}{t} \, dt$:
\begin{equation}
I_{Coh}(x)= \left| \frac{1}{\pi}\left[\text{Si}(\theta_x^+) + \text{Si}(\theta_x^-)\right]  \right|^2 \end{equation}
\begin{equation}
I_{Inc}(x)=  \frac{1}{\pi}\left[\text{Si}(2\theta_x^+) + \text{Si}(2\theta_x^-) - \frac{\sin^2\theta_x^-}{\theta_x^-} - \frac{\sin^2\theta_x^+}{\theta_x^+}\right] 
\end{equation}

\begin{figure}
\includegraphics{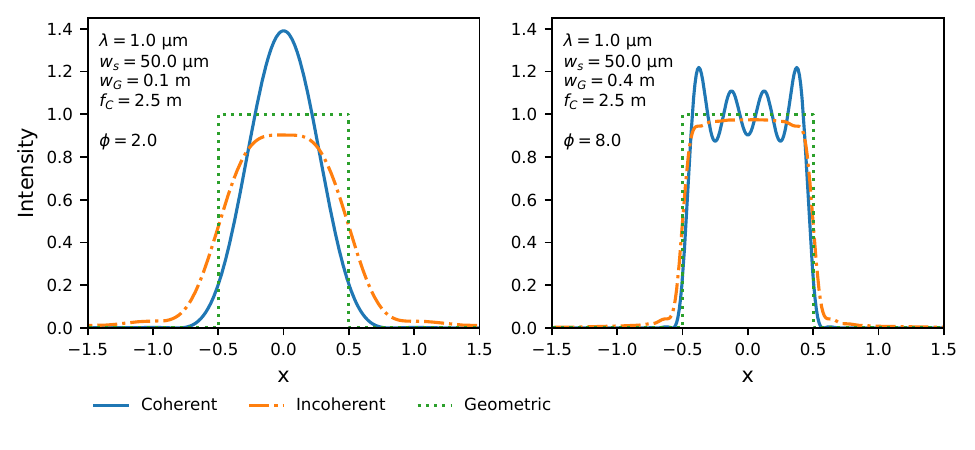}
\caption{Line profiles for the completely coherent and incoherent cases. All parameters of the spectrograph are the same in both cases except for the width of the grating aperture. The geometric slit image is shown for comparison}
\label{fig:casiniLSFs}
\end{figure}

\begin{figure}
\includegraphics{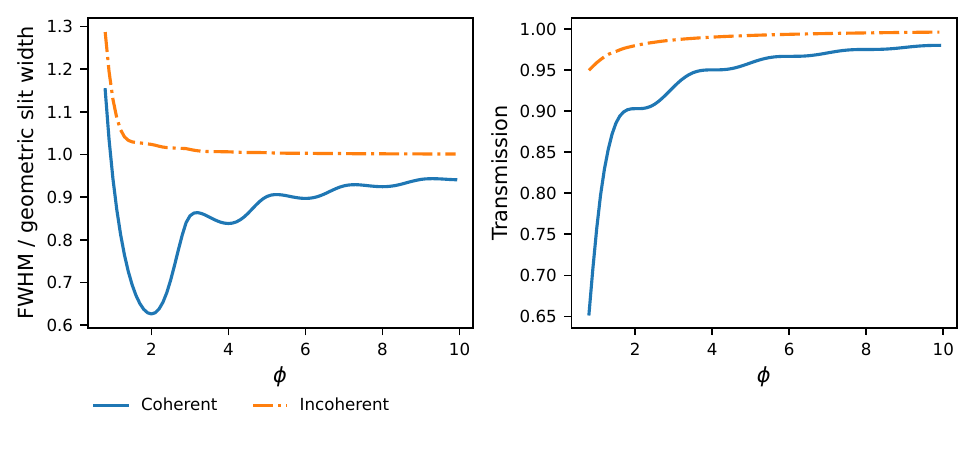}
\caption{FWHM as a function of $\phi$ for the coherent and incoherent cases discussed in the text. The transmission shows the proportion of light transmitted by the grating aperture.}
\label{fig:casiniFWHM}
\end{figure}

Examples of the resulting line profiles are shown in figure \ref{fig:casiniLSFs}. These functions were numerically evaluated to measure the full width half maximum (FWHM). This was measured from the calculated line profiles through linear interpolation and expressed as a multiple of the geometric slit image width $w_S f_L/f_C$. The FWHM can be plotted as a function of the dimensionless parameter $\phi$, as shown in figure \ref{fig:casiniFWHM}, where $\phi$ is defined as
\begin{equation}
    \phi= \frac{w_sw_G}{\lambda f_C}.
\end{equation}

We can use this plot to understand the dependence on various parameters contained within $\phi$, whether that is choosing the appropriate grating aperture size for our design, or understanding the wavelength dependent FWHM when the design parameters of the spectrograph remain constant.  The minimum FWHM in the coherent case occurs at $\phi = 2$. At this point the full width of the grating aperture is $w_G=2\lambda f_C/w_s$, matching the interval between the first zeroes in the illumination pattern in the grating plane: a sinc function with the first zeroes occurring at $\pm \lambda f_C/w_s$. The combination of slit and grating aperture is acting as a spatial filter, producing a close approximation to a Gaussian beam as can be seen in the left hand plot of figure \ref{fig:casiniLSFs}. 

\subsection{Partial coherence}

The line spread function in the coherent case is always narrower than in the incoherent case, potentially by up to 40\%, but most realistic spectrographs for astronomy will operate between these two regimes. The average degree of coherence of the slit was derived by Mielenz \cite{Mielenz:67}. This derivation is shown to be valid whether the source is imaged onto the slit (as in an astronomical spectrograph) or onto the aperture plane of the spectroscope. We can rewrite these equations in terms which are more directly related to an astronomical slit or slicing spectrograph, defining the parameter $\alpha$ as a dimensionless term giving the ratio between the slit width $w$ and the size of a diffraction limited spot at the slit, $f_{tel}\lambda/D_{tel}$. This can be directly linked to $\phi$ if we define the factor $\Omega_G$ as the oversizing of the rectangular grating aperture relative to the diameter of the geometric pupil image at the grating, so that $w_G=\Omega_GD_{tel}f_C/f_{tel}$. 

\begin{equation}
    \alpha = \frac{w_sD_{tel}}{f_{tel}\lambda}=\frac{\phi}{\Omega_G}
\end{equation}
The degree of coherence can then be written as 
\begin{equation}
    \mu_{RMS} = \sqrt{\frac{1}{\pi\alpha}\left[\text{Si}(2\pi\alpha) - \frac{\sin^2{\pi\alpha}}{\pi\alpha}\right]}.
\end{equation}

\begin{figure}
\includegraphics[center]{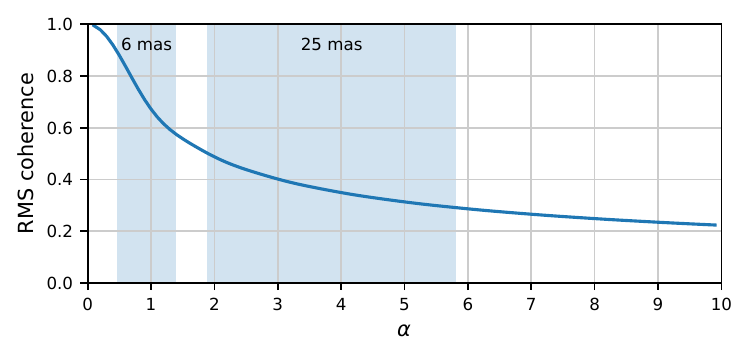}
\caption{RMS coherence as a function of $\alpha$. The two shaded regions indicate the range of $\alpha$ for the two spatial scales of HARMONI: 6 mas and 25 mas per spatial resolution element.}
\label{fig:rmscoherence}
\end{figure}

As we would expect, the degree of coherence tends to 1 when $w \ll f_{tel}\lambda/D_{tel}$, and tends to zero when $w \gg f_{tel}\lambda/D_{tel}$, however a significant degree of coherence remains even when we might consider the slit width to be relatively large, with $\mu_{RMS}$ still above 0.2 when the slit width is matched to $10\lambda/D_{tel}$. HARMONI operates in two different image scales, with slice widths on sky of 6~mas and 25~mas and across a relatively large wavelength range. As can be seen in figure \ref{fig:rmscoherence}, there is a significant degree of coherence in all of these operating conditions. At long wavelengths in the 6~mas scale we are close to the fully coherent regime, but at shorter wavelengths, or in the 25 mas scale, we cannot make the simplification of assuming either the fully coherent or fully incoherent case.  

\subsection{Modelling the partially coherent line profile}

In order to include the partial coherence effects in the calculated line profile we must start from the entrance pupil plane before the slit aperture. Mielenz gives an expression for the complex amplitude in the image plane due to a point source in the entrance pupil plane at $q\xi$ in terms of the Sine and Cosine integral functions:

\begin{equation}
    U(v', v_0,q\xi)=\frac{1}{2}(V+iW)e^{-iv'q\xi}
\end{equation}
where
\[V = \textrm{Si}((1+q\xi)(v'+v_0)) + \textrm{Si}((1-q\xi)(v'+v_0)) - \textrm{Si}((1+q\xi)(v'-v_0)) - \textrm{Si}((1-q\xi)(v'-v_0))\]
\[W = -\textrm{Ci}|(1+q\xi)(v'+v_0)| + \textrm{Ci}|(1-q\xi)(v'+v_0)| + \textrm{Ci}|(1+q\xi)(v'-v_0)| - \textrm{Ci}|(1-q\xi)(v'-v_0)|\]

Relating this to the terms defined above, $q = 1/\Omega_G$, $v_0 = \pi\phi/2$, $v'=\pi x' w_G/f_L\lambda$ where $x'$ is the distance from the line centre in the output focal plane. 

The partially coherent intensity profile can be obtained by integrating this transfer function over the entrance pupil. We introduce an additional apodising factor here to the expression given by Mielenz to account for the effect of a circular, rather than rectangular, entrance pupil aperture.

\begin{equation}
    I(v')=\int^{+1}_{-1}\sqrt{1-\xi^2}\ |U(v', v_0,q\xi)|^2\,d\xi
\end{equation}

\begin{figure}
\includegraphics[center]{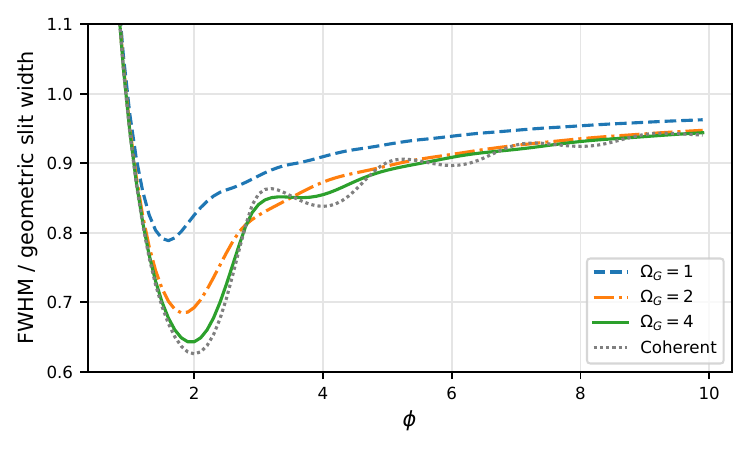}
\caption{The FWHM of line profiles generated using the expressions for the partial coherence case described in the text.}
\label{fig:mielenzFWHM}
\end{figure}

This function was evaluated by numerical integration using a Python script. The FWHM of each profile was measured directly from the oversampled line profile data to generate the data plotted in figure \ref{fig:mielenzFWHM}. This plot shows the expected FWHM as a function of $\phi$ for three different cases relevant to the design of a realistic spectrometer, with $\Omega_G = 1$ describing the situation where the grating aperture is sized to match the geometric pupil diameter, and $\Omega_G = 2$ and 4 showing the effect of oversizing the grating aperture. 

\subsection{Application to HARMONI}

While HARMONI is a complex instrument, for the purposes of this initial analysis we can represent it by a simple paraxial model as shown in figure \ref{fig:paraxial model}. The first paraxial lens now includes the combined effects of the ELT and MORFEO as well as the pre-optics and IFU within HARMONI. Since we are modelling the spectral line resulting from a spatially uniform extended source (e.g. a calibration arc line or a sky emission line) we are insensitive to any optical aberrations before the slit plane. Using values of $w_s= \SI{130}{\um}$, $w_G= 50\,$mm and $f_C = 1319$ mm we find that the term $\phi$ defined above varies from $\phi = 6.2$ at \SI{0.8}{\um} to $\phi = 2.0$ at \SI{2.45}{\um}. When operating in the 25 mas spatial scale the geometric spatial ratio at the slit is matched to the grating width with a small margin. This corresponds to a grating oversizing of $\Omega_G= 1.06$. Switching to the finer 6 mas scale produces a slower optical beam at the slit as the image is magnified onto the slit plane. This reduces the diameter of the entrance pupil so the grating aperture is now oversized relative to the geometric pupil image with $\Omega_G= 4.4$. Looking at the curves plotted in figure \ref{fig:mielenzFWHM} we see that for long wavelengths in HARMONI this places us at $\phi \approx2$ where the discrepency is largest between the coherent and geometric slit image widths, and where the sensitivity to $\Omega_G$ is greatest.

\begin{figure}
\includegraphics[center]{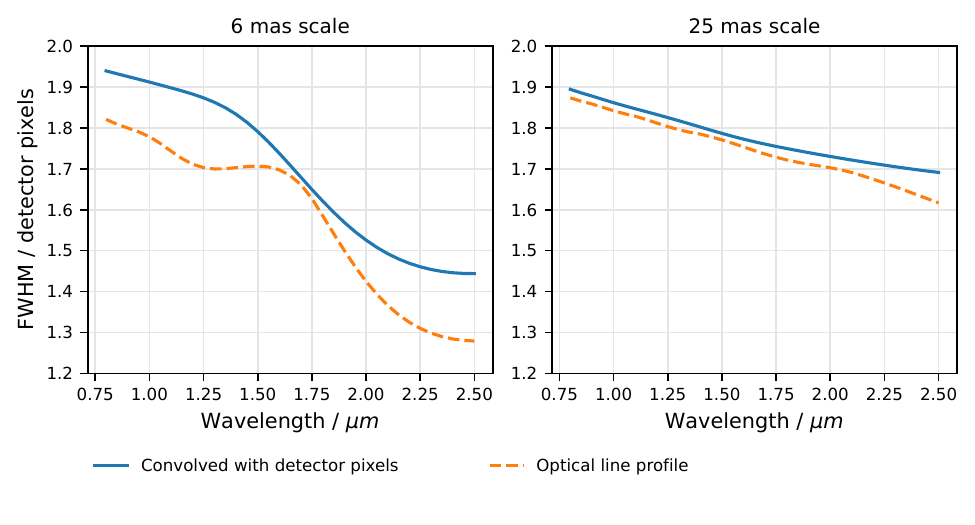}
\caption{The FWHM of line profiles in an idealised model of the HARMONI spectrograph in the two spatial scales (6 mas per spaxel and 25 mas per spaxel). Dashed lines show the optical line profile while the solid line includes the effect of convolution with a \SI{15}{\um} pixel.}
\label{fig:harmoniFWHM}
\end{figure}

The FWHM values are plotted as a function of wavelength for these HARMONI parameters in figure \ref{fig:harmoniFWHM}. The magnification from slit to detector is 0.231 so the geometric slit image width at the detector is \SI{30}{\um}, matching 2 pixels on the HAWAII-4RG detector. The pixel sampling effect is included in this plot by convolving the optical line profile with a \SI{15}{\um} wide boxcar function before the FWHM is measured.

\begin{figure}
\includegraphics[center]{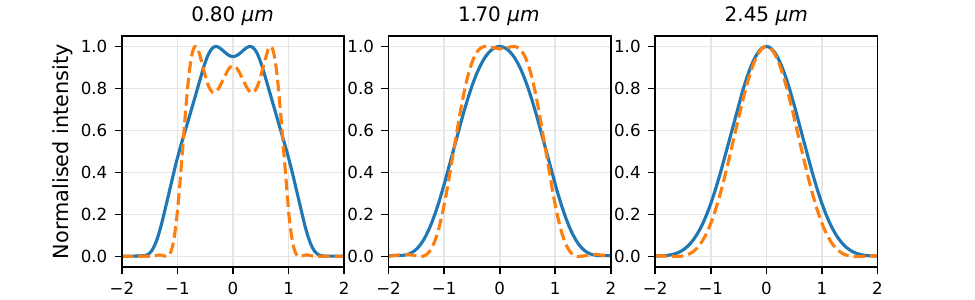}
\caption{Line profiles in an idealised model of the HARMONI spectrograph at three selected wavelengths in the 6 mas spaxel scale mode. Dashed lines show the optical line profile while the solid line includes the effect of convolution with a \SI{15}{\um} pixel.}
\label{fig:harmoniprofiles}
\end{figure}

From these plots we see that when we take diffraction and detector pixel sampling into account, we expect the spectral lines observed in HARMONI to have a FWHM less than 2 pixels at all wavelengths in both the 6 mas and 25 mas scales, resulting in undersampling of the spectrum. The undersampling is most pronounced at long wavelengths in the 6 mas scale, where diffraction effects dominate. It is interesting to note the effect of convolving with the pixel function, and how this varies with wavelength in the 6 mas scale due to the changing shape of the optical line profile as shown in figure \ref{fig:harmoniprofiles}. At the shortest wavelengths the optical line profile of the 6 mas scale is narrower than the same wavelength in the 25 mas scale, but once the detector pixels are included the measured width is slightly higher in the 6 mas scale than in the 25 mas scale.

\section{Mitigation of expected undersampling in HARMONI}

The predicted undersampling of spectra from HARMONI would have a significant negative impact on the performance of the instrument, as discussed above. A trade study was carried out to evaluate a wide range of potential approaches to mitigate this undersampling issue. There are a number of constraints to be considered. The HARMONI design is already well developed. In particular, the integral field unit optics are already in manufacture. Any mitigation options that involve redesign of the integral field unit would result in significant extra costs and delays to the project. The project has recently been through a rescope process to simplify the instrument and reduce technical risk. One of the key aims was to minimise the number of cryogenic mechanisms, so it is also undesirable to follow any mitigation approach that requires the introduction of additional mechanisms. It is also challenging to find a mitigation approach which achieves 2 pixel sampling across the full wavelength range for the 6 mas scale without introducing excessive oversampling in the 25 mas scale or at shorter wavelengths, with a corresponding degradation of spectral resolution.

We have adopted a mitigation approach using the grating geometry to introduce a small anamorphic magnification factor. Each VPH grating is sandwiched between a pair of prisms to form a grism. This gives additional degrees of freedom to the design, allowing different anamorphic factors to be introduced for each wavelength band. There are still compromises to be made: using the angle of the grism allows the anamorphic factor to be set at the centre of each band, but the variation across the band is essentially fixed. At longer wavelengths we need a mitigation factor which achieves an acceptable sampling in the 6 mas scale while not excessively degrading the spectral resolution in the 25 mas scale.

\begin{figure}
\includegraphics[width=\textwidth]{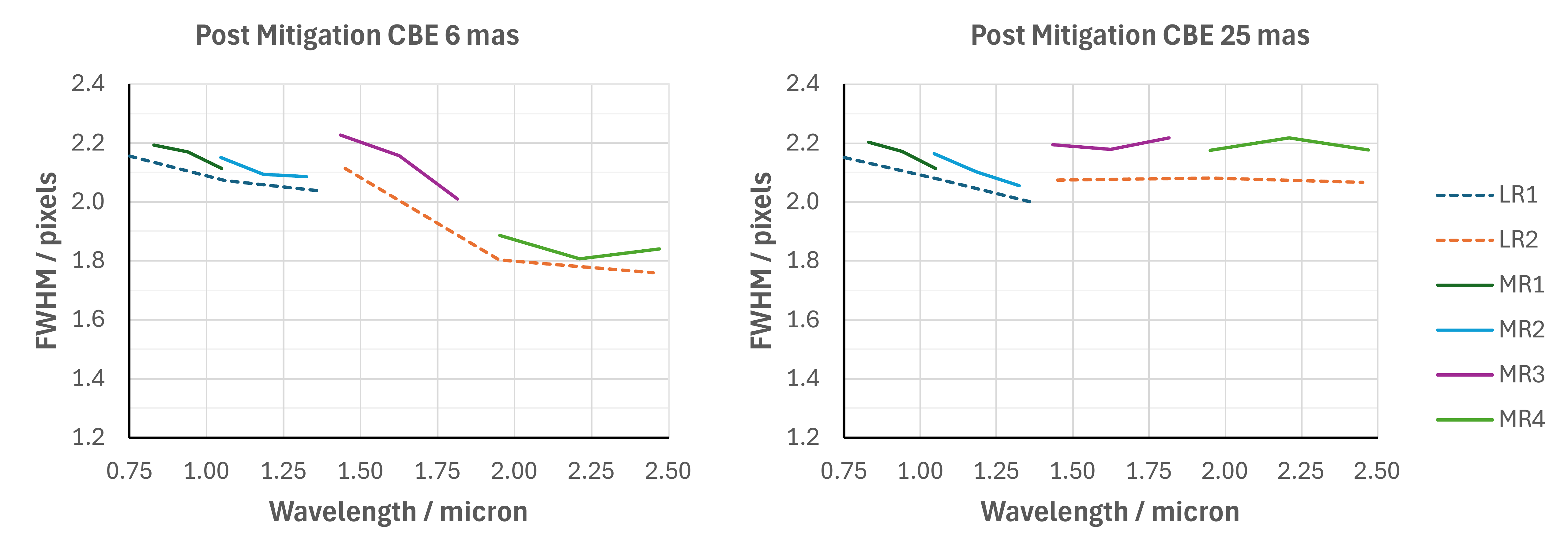}
\caption{The current best estimate for the HARMONI spectral FWHM with the proposed mitigation using the grism geometry implemented as discussed in the text.}
\label{fig:mitigated_FWHM}
\end{figure}

The calculations shown so far are all based on an ideal spectrograph. In reality the spectral lines will be broadened by other effects including optical aberrations, grating errors, vibration and detector effects. If we base our correction factors on the ideal model then there is a risk of over-correction, resulting in degradation of the spectral resolution. Modelling the combined effect of optical aberrations and diffraction is not straightforward and is discussed further below. In order to allow design work to continue on the spectrograph a spreadsheet based error budgeting approach was used to estimate the impact of these effects. The contribution of these effects to the FWHM were combined on an RSS basis to give an estimate of the as-built FWHM and an associated uncertainty range. The anamorphic factor $f_j$ required to achieve 2 pixel sampling was derived from the FWHM values for each wavelength range. Limits were applied so that $1 \le f_j \le 1.15$ to reflect the range of grating geometries that can practically be achieved within the constraints of the spectrograph design. These parameters are still to be finalised but the current best estimate of the FWHM values with this mitigation in place are shown in figure \ref{fig:mitigated_FWHM}. As we can see here, sampling of ~2 pixels is achieved at all wavelengths in the 25 mas scale. In the 6 mas scale we still expect a small degree of undersampling at longer wavelengths, which is judged to provide an acceptable compromise between sampling in the 6 mas scale and resolution in the 25 mas scale.

\section{Modelling a realistic spectrograph}

To model the real expected performance of the system we need to include geometric aberrations as well as the diffraction effects discussed above. Combining these in a single model is very challenging when we have significant diffraction effects and beam clipping occuring in both the pupil plane and the image plane. The method adopted so far within the HARMONI project is to simulate the extended illumination not by integrating over the entrance pupil as discussed above, but instead to scan a diffraction limited point source across the entrance slit and integrate the intensity profiles at the detector. This is discussed in more detail in a previous paper\cite{Todd2024b}.

The case of a monochromatic diffraction limited point source can be analysed using a Fourier optics approach. The input pupil is uniformly illuminated by a coherent plane wave. The light can be propagated through the system using a series of Fourier transforms with clipping applied at the slit and the pupil planes to represent the apertures in the system. A simple model was created using the PROPER optical propagation library and a series of Python scripts. The aberrations of the optical model can be included by extracting a wavefront map from the Zemax Opticstudio model of the system for given field point and wavelength, and inserting this into the Fourier optics model as a phase error in the pupil plane. 

This approach is computationally slow, and requires multiple wavefront maps to be generated and exported from Zemax Opticstudio. No simpler solution, allowing the use of a single optical model, has been identified at this point. The ERIS line profiles were successfully modelled using the Code V Partial Coherence tool \cite{George2017}. This may be a potential approach but the current HARMONI spectrograph models cannot be directly imported into Code V due to the freeform surface types used in the optical design of the spectrograph, which cannot be directly translated into Code V surface types, so additional work-arounds would be needed in order to use this tool. 

\section{Conclusions}

When spectrographs are designed to take advantage of near-diffraction limited resolution on AO corrected telescopes, or in space, the line spread function will be dominated by diffraction effects. In many cases, as shown here, this will result in a FWHM significantly less that the geometric slit width. These effects have long been known, but rarely considered in the initial design phase of new instruments. In this paper we have presented these models in a form that can directly be applied when calculating top level optical parameters for a new spectrograph design. The dimensionless parameters $\phi$ and $\Omega_G$ can be calculated and used to give an initial estimate of the significance of these effects as shown in figures \ref{fig:rmscoherence} and \ref{fig:mielenzFWHM}. The top level optical parameters can be adjusted to give acceptable sampling before stating detailed optical design.

In the case of HARMONI this has been identified during the design process. The design is already well developed, placing many constraints on the potential routes to mitigate this issue due to the impacts on the rest of the system. The challenge is increased because HARMONI operates across a broad range of parameter space as shown in figure \ref{fig:rmscoherence}. We have identified an approach using the geometry of grisms in the spectrograph to provide small anamorphic magnification factors which are predicted to give an acceptable level of sampling across the wavelength range in both spatial scales. 

Further work is needed to develop a robust process and a set of tools to model these diffraction effects in combination with the optical aberrations that will be present in a real spectrograph.

\acknowledgments % equivalent to \section*{ACKNOWLEDGMENTS}       
 
The authors of this paper were members of a HARMONI Working Group set up to investigate the predicted spectral undersampling and potential mitigations. We would like to thank the wider HARMONI Consortium and colleagues at ESO for their contributions to this study.

% References
\bibliographystyle{spiebib} % makes bibtex use spiebib.bst
\bibliography{spectral_sampling} % bibliography data in report.bib

\end{document}